\documentclass[12pt]{article}

\usepackage[colorlinks=true, a4paper=true, pdfstartview=FitV,
linkcolor=blue, citecolor=blue, urlcolor=blue]{hyperref}

\usepackage{mwe}
\usepackage{eurosym}
\usepackage{amssymb}
\usepackage{color}
\usepackage{amsmath}
\usepackage{graphicx} 
\usepackage{subfig}
\usepackage{hyperref}
\usepackage{makecell}
\usepackage{lineno}   

\title{Performance Indicators of Wind Energy Production}

\begin{document}

\maketitle

\begin{center}

Guglielmo D'Amico\\
Dipartimento di Farmacia, 
Universit\`a `G. D'Annunzio' di Chieti-Pescara,  66013 Chieti, Italy\\
\vskip .3cm
Filippo Petroni\footnote{Corresponding Author. {\it Email address:} {\bf fpetroni@unica.it}}\\
Dipartimento di Scienze Economiche ed Aziendali,
Universit\`a degli studi di Cagliari, 09123 Cagliari, Italy\\
\vskip .3cm
Flavio Prattico\\
Dipartimento di Ingegneria Industriale e dell'Informazione e di Economia,\\
Universit\`a degli studi dell'Aquila,\\
67100 L'Aquila, Italy\\
\end{center}

\bigskip

\begin{abstract}
Modeling wind speed is one of the key element when dealing with the production of energy through wind turbines.  A good model can be used for forecasting, site evaluation, turbines design and many other purposes. In this work we are interested in the analysis of the future financial cash flows generated by selling the electrical energy produced. We apply an indexed semi-Markov model of wind speed that has been shown, in previous investigation, to reproduce accurately the statistical behavior of wind speed. The model is applied to the evaluation of financial indicators like the Internal Rate of Return, semi-Elasticity and relative Convexity that are widely used for the assessment of the profitability of an investment and for the measurement and analysis of interest rate risk. We compare the computation of these indicators for real and synthetic data. Moreover, we propose a new indicator that can be used to compare the degree of utilization of different power plants.
\end{abstract}

\vskip 1cm



\section{Introduction}

Renewable energy is becoming a big player in the world energy production. Already in 2009, the world generating capacity through renewable energy was of 26\% against the 66\% coming from fossil fuel \cite{fouquet2013policy}. In this scenario the wind energy has been growing continuously \cite{Suom14} although the intrinsic volatility of wind speed requires accurate evaluation of wind speed data before installing wind turbines. Due to its random behavior \cite{sahin2012generation} and the intrinsic difficulty of storage \cite{johnston2015methodology}, many researchers are still working on the quantification of available energy and performances estimation of wind projects \cite{chang2011estimation,chang2011performance,nagai2009performance,danao2013experimental}.

For this reason the wind speed modeling play an important role in the electrical power generation from wind speed. \\
\indent Accurate models permit the forecast of future wind speed and energy production (see for example \cite{jiang2013very,liu2012comparison,cassola2012wind}.  Many scholars have proposed new models that can allow the prediction of wind speed, minutes, hours or days ahead. Many of these models are based on neural networks,  autoregressive models, Markov chains \cite{Song14}, hybrid models where the previous mentioned models are combined, see \cite{review} and the references therein. Often, these models are either focused on specific time scale forecasting, or synthetic time series generation. 
In \cite{wind1,wind3,wind4}, different semi-Markov models were applied to wind speed modeling and it was shown that the semi-Markov framework over perform the Markov models and therefore they should be preferred in the modeling of wind speed. 

The approach we propose here is based on an indexed semi-Markov chain (ISMC) model that was advanced in \cite{wind2} and applied to the generation of synthetic wind speed time series. In \cite{wind2} we showed that the ISMC model is able to reproduce correctly the statistical behavior of wind speed. The ISMC model is a non-parametric model because it does not require any assumption on the form of the distribution function of wind speed.
Furthermore, it is able to forecast wind speed at different time scale without loosing the goodness of forecasting which is almost independent from the time horizon \cite{wind5}. 

The main purpose of this paper is to present a profitability analysis of wind energy production through economical indexes. Indeed, we first present a comparison between the real energy production and that obtained by implementing the ISMC model on real data. Then we advance a new indicator, that can be interpret as a degree of utilization of the wind turbine, and we compute it for real and synthetic data. Finally we consider financial indicators of the cash flow generated by selling the electrical energy produced in the wind farm. The cash flow sequences are unknown because they depend on the future random sequences of wind speed that are generated according to the ISMC model. The cash flows are subject to the interest rate risk which is an important factor that affects the risky operation of producing wind energy and the financial indicators gives the possibility to manage such a risk.\\
\indent The paper is divided in this way: first, the database and a commercial wind turbine  is illustrated. Second, the ISMC model is shortly described as the theoretical support of the empirical application. Third, a real data application is executed involving the proposal of a new index and the computation of classical financial indicators in our stochastic framework.

\section{Database and commercial wind turbine}
The wind speed data that we employ for the analysis are the same used in precedent works \cite{wind1,wind3,wind4}.  
They cover almost five year with time sampling of 10 minutes and the measurement instruments are located at 22 $m$ above the ground. Since we need to test the model for the energy production of a commercial wind turbine, we transpose our database at the turbine rotor altitude using a well known relation, see e.g. \cite{gass}:
\begin{equation}\label{height}
v_h = v_{rif} \left( \frac{h}{h_{rif}} \right)^{\alpha} \;\;\;\;\; \alpha= \frac{1}{ln\frac{h}{z_0}}
\end{equation}
where $v_h$ is the wind speed at the height $h$ of the wind turbine hub and $v_{rif}$ is the value of the wind speed at the height $h_{rif}$ of the instrument. In our application $h=50m$ and $h_{rif}=22m$. The symbol $z_0$ denotes a parameter that takes into account the morphology of the area near the wind turbine. For a region without buildings or trees it varies from 0.01 to 0.001, instead for the offshore application it is equal to 0.0001. For our analysis we consider a mean value for an onshore application, then we choose $z_0 =0.005$.



\begin{figure}
\centering
\includegraphics[height=8cm]{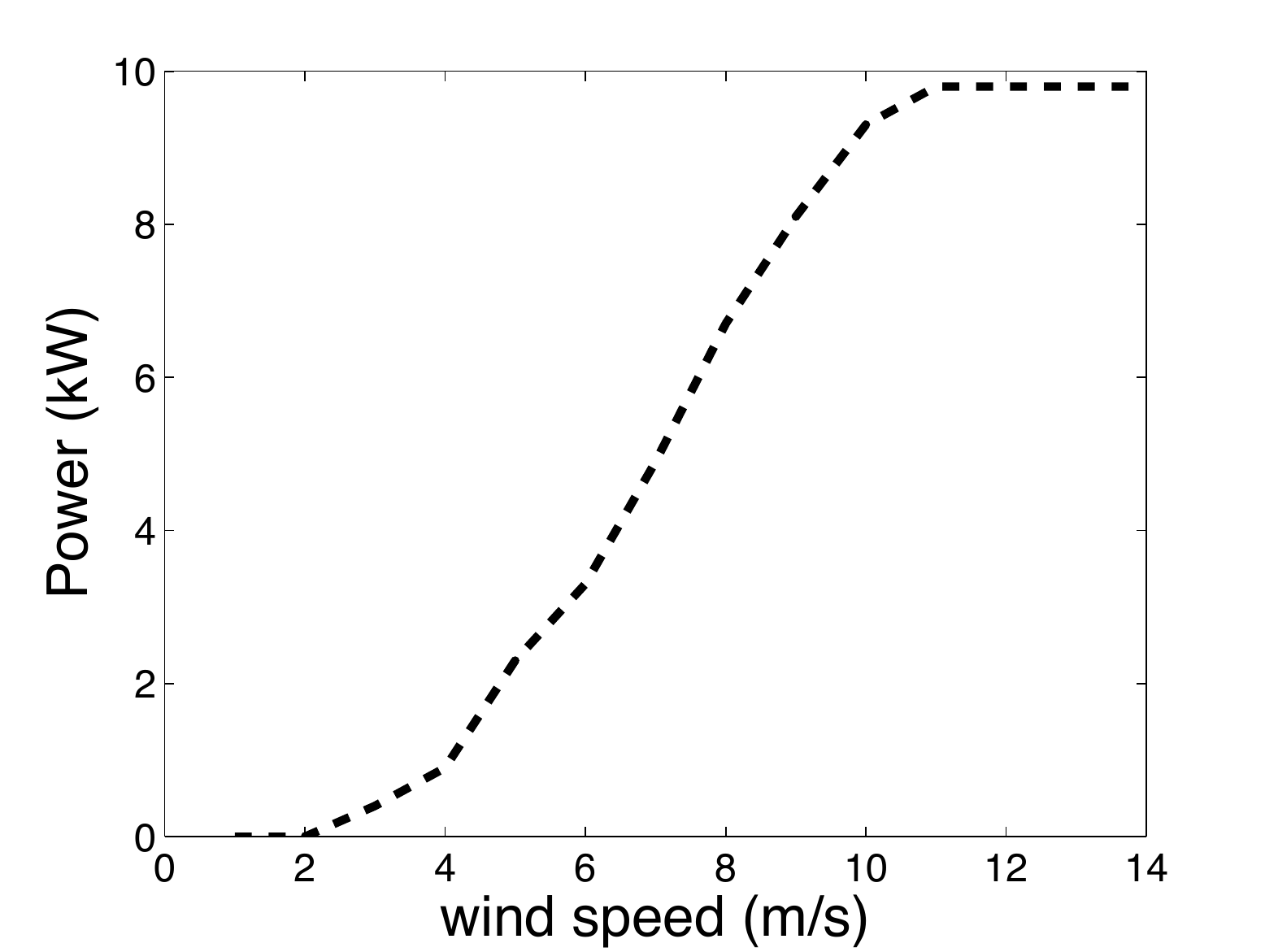}
\caption{Power curve of the 10 kW Aircon wind turbine}\label{pc1}
\end{figure}

The ISMC model is based on a discrete state space, for this reason we select 8 wind speed values, see Table \ref{st}, chosen to cover all the wind speed distribution. Table \ref{st} shows the wind speed states with their related wind speed ranges.
\begin{table}
\begin{center}
\begin{tabular}{|c|*{2}{c|}|}
     \hline
Sate & Wind speed range $m/s$  \\ \hline
1 & 0 to 3  \\ \hline
2 & 3 - 4  \\ \hline
3 & 4 - 5  \\ \hline
4 & 5 - 6 \\ \hline
5 & 6 - 7  \\ \hline
6 & 7 - 8 \\ \hline
7 & 8 - 9  \\ \hline
8 & $>$9  \\ \hline
\end{tabular} 
\caption{Wind speed discretization}
\label{st} 
\end{center}
\end{table}
The transformation between wind speed and produced energy is made through a commercial 10 kW Aircon HAWT wind turbine. The power curve of this turbine is plotted in Figure \ref{pc1} and its discrete numerical values are showed in Table \ref{pc}. The power curve of a specific wind turbine represents the energy production as a function of the wind speed. The continuous values of the power curve are obtained with linear interpolation of the discretized values.

\begin{table}
\begin{center}
\begin{tabular}{|c|*{2}{c|}|}
     \hline
Wind speed [m/s] & Power [kW]  \\ \hline
1 & 0  \\ \hline
2 & 0  \\ \hline
2.5 & 0.4  \\ \hline
4 & 0.9 \\ \hline
5 & 2.3  \\ \hline
6 & 3.3  \\ \hline
7 & 4.9  \\ \hline
8 & 6.7  \\ \hline
9 & 8.1  \\ \hline
10 & 9.3  \\ \hline
11 & 9.8  \\ \hline
11.5 to 25 & 9.8  \\ \hline
\end{tabular} 
\caption{Discrete values of the power curve of the 10 kW Aircon wind turbine}
\label{pc} 
\end{center}
\end{table}

\section{The indexed semi-Markov chain model}
The general formulation of the ISMC model has been developed in references \cite{wind2,wind3,wind4}, here we only discuss it informally.

Semi-Markov processes have similar idea as those that generate Markov processes. The processes are both described by a set of finite states $v_n$ whose transitions are ruled by a transition probability matrix. The semi-Markov process differs from the Markov process because the transition times $T_n$ are generated according to random variables. Indeed, the time between transitions $T_{n+1}-T_n$ is random and may be modeled by means of any type of distribution functions.
In studies concerning wind speed modeling the states $v_n$ indicates discretized wind speed at the nth transition and $T_n$ the time in which the nth change of wind speed occurs. 

In order to better represent the statistical characteristics of wind speed, the idea of an ISMC model was advanced in the field of wind speed, see \cite{wind2}. 
The novelty, with respect to the semi-Markov case, consists in the introduction of a third random variable defined as follow:
\begin{equation}
U_{n}^{m}= \sum_{k=0}^{m} v_{n-1-k} \cdot \frac{T_{n-k}-T_{n-1-k}}{T_{n}-T_{n-1-m}}. 
\end{equation}
This variable can be interpreted  as a moving average of order $m+1$ executed on the series of the past wind speed values $(v_{n-1-k})$ with weights given by the fractions of sojourn times in that wind speed $(T_{n-k}-T_{n-1-k})$ with respect to the interval time on which the average is executed $(T_n-T_{n-1-m})$.\\
\indent The ISMC model considers that the probability of changes in wind speed do depend also on this new variable. In a simple semi-Markov model it depends only on the present wind speed and on the transition time. The ISMC model can be considered as an $m$ order semi-Markov chain where the dependence is given only by the averaged last $m$ states.
Also the process $U^m$ has been discretized, Table \ref{Um} shows the states of the process and their values.
\begin{table}
\begin{center}
\begin{tabular}{|c|*{2}{c|}|}
     \hline
Sate & $U^m$ range $m/s$  \\ \hline
1 & 0 to 2.1  \\ \hline
2 & 2.1 - 2.6  \\ \hline
3 & 2.6 - 3.4  \\ \hline
4 & 3.4 - 6 \\ \hline
5 & $>$6  \\ \hline
\end{tabular} 
\caption{$U^m$ processes discretization}
\label{Um} 
\end{center}
\end{table}

The parameter $m$ must be optimized as a function of the specific database. The optimization is made by finding the value of $m$ that realizes the minimum of the root mean square error (RMSE) between the autocorrelation functions (ACF) of real and simulated data, see \cite{wind2}. In our analysis $m=7$.
The ISMC model revealed to be particularly efficient in reproducing together the probability density function of wind speed and the autocorrelation function, see \cite{wind2}.

The one step transition probability matrix can be evaluated by considering the counting transition between the three random variables considered before. Then, the probability $p_{i,j}(t,u)$ represents the transition probability from the actual wind speed state $i$, to the wind speed state $j$, given that the sojourn time spent in the state $i$ is equal to $t$ and the value of the process $U^m$ is $u$. These probabilities can be computed as:

\begin{equation}
p_{i,j}(t,u)= \frac{ n_{i,j} (t,u) }{\sum\limits_{j} n_{i,j}(t,u)},
\end{equation}
\label{pri}

\noindent where $n_{ij}(t,u)$ is the total number of transitions observed in the database from state $i$ to state $j$ in next period having a sojourn time spent in the wind speed $i$ equal to $t$ and the value of the index process  equal to $u$.

\section{Economic application}
Mathematical models of wind speed can be used to assess the economic validity of an investment in a wind farm. In this section, we show how the ISMC model can be used to evaluate useful economic indicators like the Internal Rate of Return, the semi-Elasticity and the relative Convexity of the cash flow generated by the investment on a wind farm in a specific site. We conduct two separate analysis: in the first one we consider an investment for a period of 15 years, which corresponds to the period in which the government incentives on the energy price in Italy are still present. In the second case we consider an investment without the government incentives and in this case, being the price of the energy lower than in the previous case, the period of the investment is settled to be greater and equal to 30 years. \\
\indent  Since the database of wind speed is composed of only 5 years of wind speed measures, we apply a bootstrap procedure in order to have the sufficient number of data for the analysis. The database has been adapted to the blade height by using equation \ref{height}.
To compare real data with the ISMC model we generate, through Monte Carlo simulation, 1000 possible scenarios of wind speed time series of length 15 years for the first case and 30 years for the second case. The sampling period is 10 minutes.

\subsection{Energy production}
\begin{figure}
\centering
\includegraphics[height=7cm]{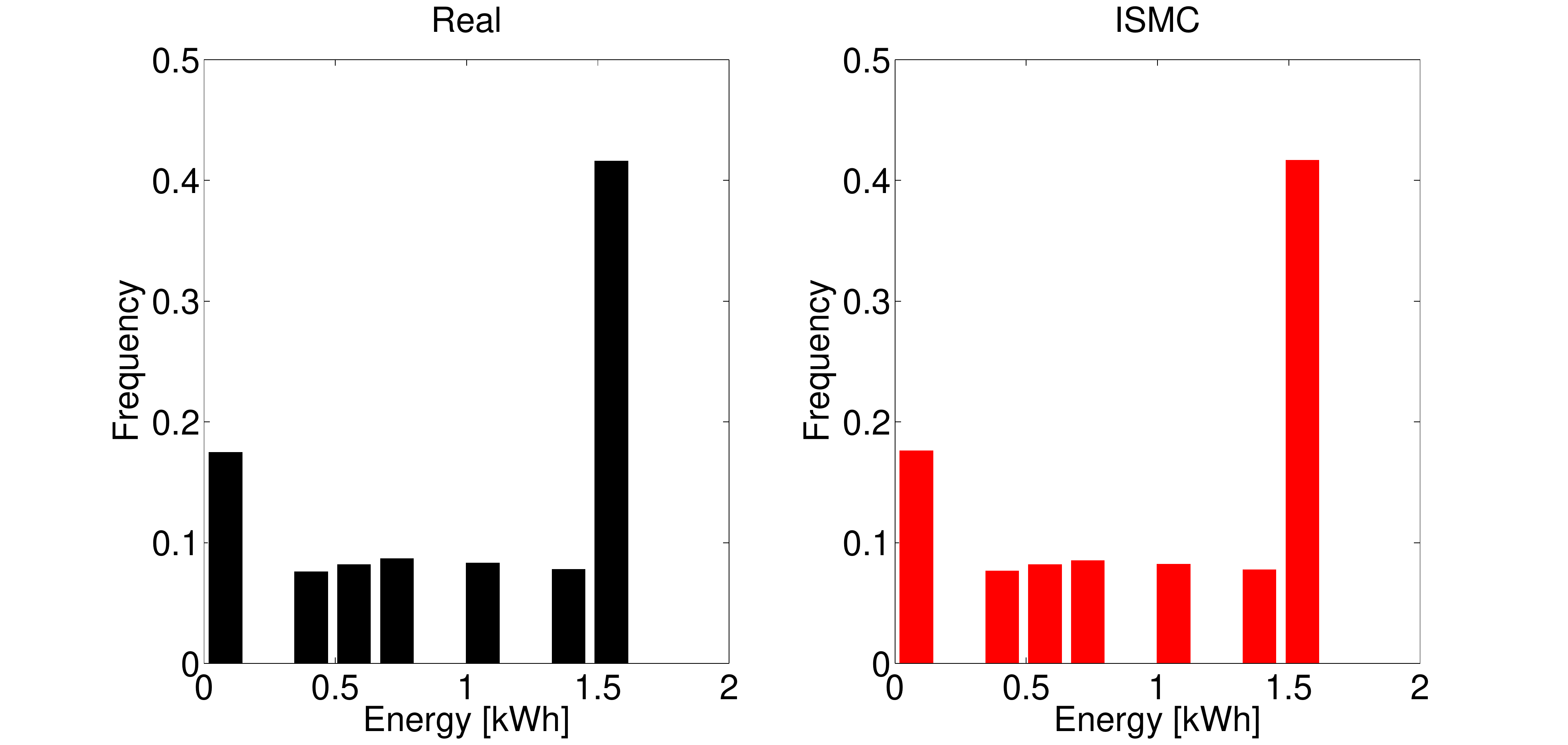}
\caption{Histogram of the electrical energy produced in 10 minutes. Comparison between real and simulated data.}\label{enr1}
\end{figure}

According to the power curve described in Figure \ref{pc1} and Table \ref{pc} wind speed has been transformed into power and then in produced energy in a specific time window.
In Figure \ref{enr1} it is shown the comparison between the distribution of the produced energy in a 10 minutes interval for real and synthetic data. Although the transformation from wind speed into energy is nonlinear, the ISMC model is able to reproduce correctly also a real case of energy production. In fact the values obtained in the two cases are almost identical. The particular form of these distributions is due to the existence of a cut-in wind speed, under which no energy is produced, and a part of the power curve that remains with constant value, from 10 $m/s$ until 32 $m/s$, in which the wind turbine produce energy at its rated power. This explain the existence of the two modes.

\begin{table}
	\begin{center}
		\begin{tabular}{lccccccc}
			\hline
			Annual Energy        &        &        &          &      &  & &  \\
			\hline
			& Mean   & STD    & Skewness & Kurtosis & JB & jB stat & p-value \\
			Real   		& 147.4  & 42.90  & -0.331   & 2.70   &  Rejected & 29 & 0.001\\
			Synthetic   & 150.5  & 37.82  & -0.283   & 2.65  &  Rejected & 24 &0.001 \\
			\hline
		\end{tabular}
		\caption{Mean value, standard deviation, Skewness, Kurtosis and JB test for real and synthetic energy produced}
		\label{anen} 
	\end{center}
\end{table}

\begin{figure}
\centering
\includegraphics[height=7cm]{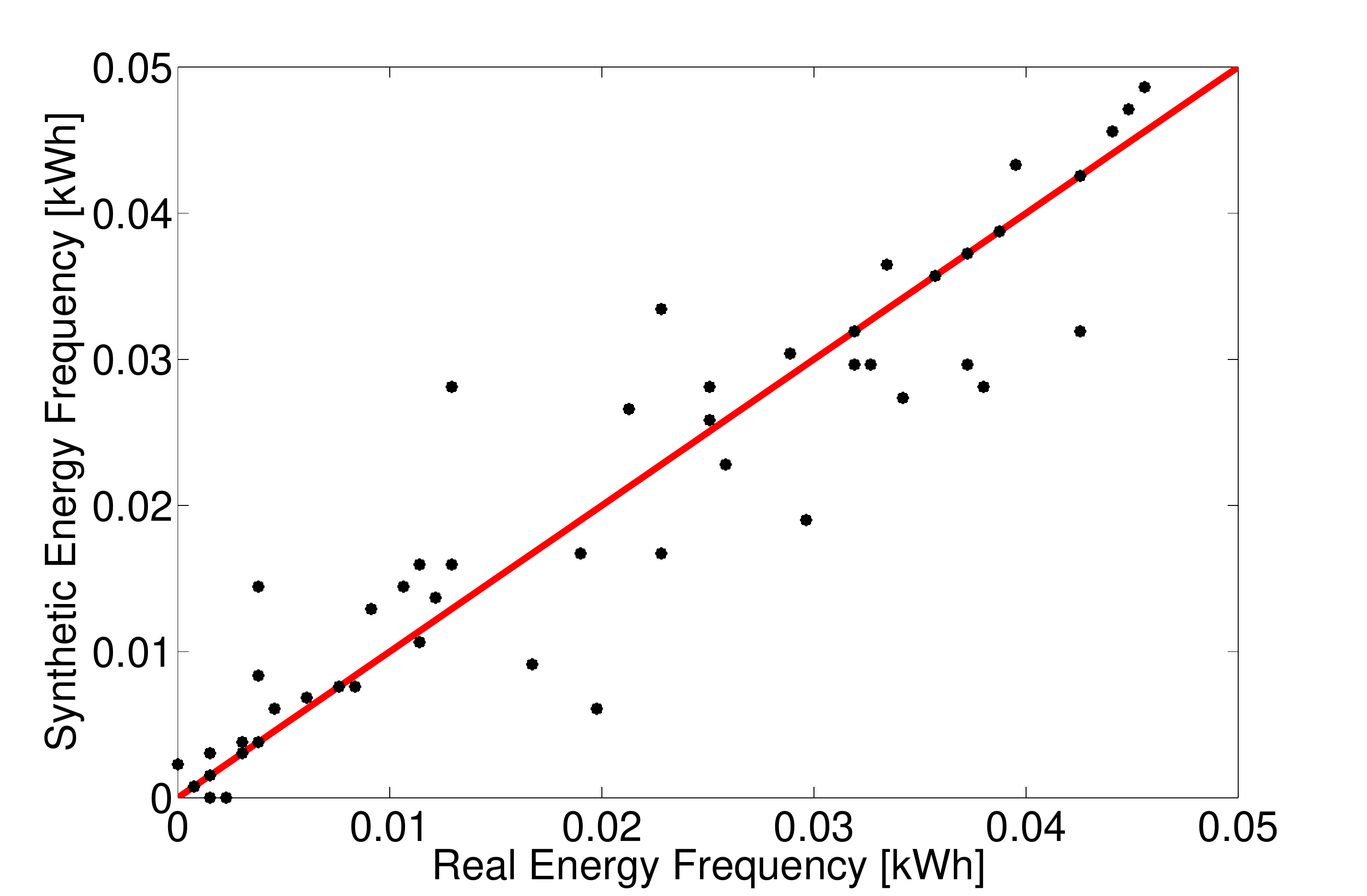}
\caption{Synthetic annual energy distribution as a function of real annual energy distribution.}\label{enr2}
\end{figure}

In table \ref{anen} we give some statistics of the annual energy distribution for real and synthetic data. We included the results for the Jarque–Bera test (JB) which is a goodness-of-fit test of whether sample data have the skewness and kurtosis matching a normal distribution. The test rejects the normality hypothesis for the annual energy. This is mainly due to the absence of a Gaussian tail of the distribution on its right side (see the distribution figure in \cite{windbar}).\\
\indent In Figure \ref{enr2}, instead, we show the synthetic annual energy distribution against the real annual energy distribution. Dots close to the straight line demonstrate the appropriateness of the ISMC model. The coefficient of determination for this set of data is  $r^2=0.89$. 

\begin{figure}
\centering
\includegraphics[height=8cm]{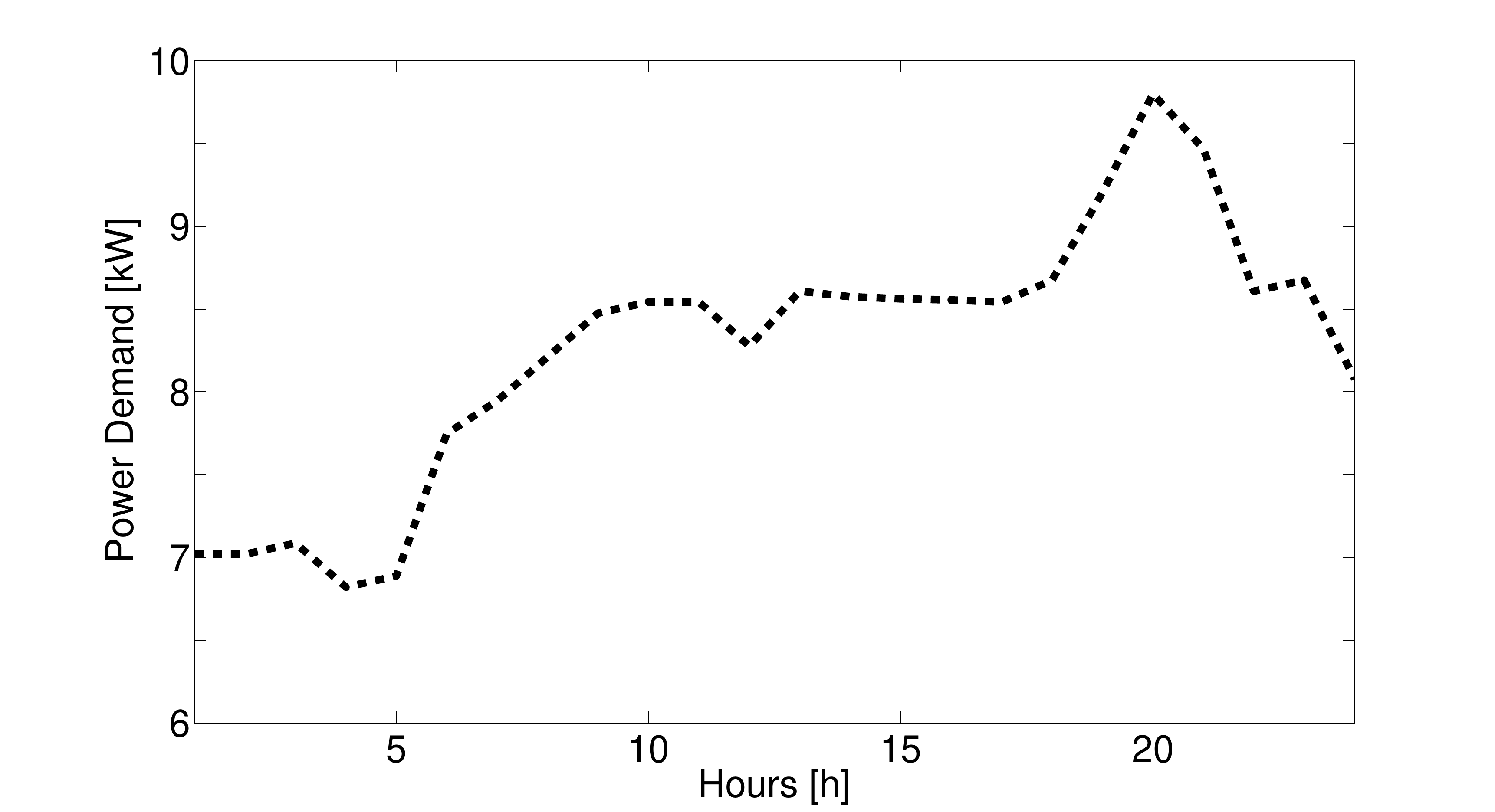}
\caption{Power demand scaled on the commercial wind turbine rated power}\label{cons}
\end{figure}

As completion of this section, we show another comparison between real and synthetic energy produced. Following a similar approach of \cite{mabel}, we propose a new indicator that quantify the utilization degree of a wind farm. We define the satisfying power demand percentage (SPDP) as the part of time in which the wind turbine is satisfying the electrical power demand related to the specific considered period. Formally it can be defined as follow:
\begin{equation}
SPDP=\frac{T(P_a - P_r \geq 0)}{T}
\end{equation} 
where the numerator represents the period in which the electrical power available ($P_a $) satisfies the electrical power required ($P_r $) and $T$ is the number of unit of time where the inequality $P_a - P_r \geq 0$ is satisfied in the considered time horizon.
To compute this indicator we evaluate the power available (practically the power available on the wind turbine) and we compare it with a scaled electrical power demand. We use the data of \cite{mabel} and scaling its maximum value. Particularly in \cite{mabel} the greatest value (during the day) of the power demand is 14.8 $MW$. We scaled all the curve to have the maximum value equal to the rated power of the chosen commercial wind turbine, then 9.8 $kW$. The result of the scaling procedure is showed in Figure \ref{cons}. This represents the power demand in a day, we extend it for the considered period (five years) without considering any seasonality.

\begin{figure}
\centering
\includegraphics[height=7cm]{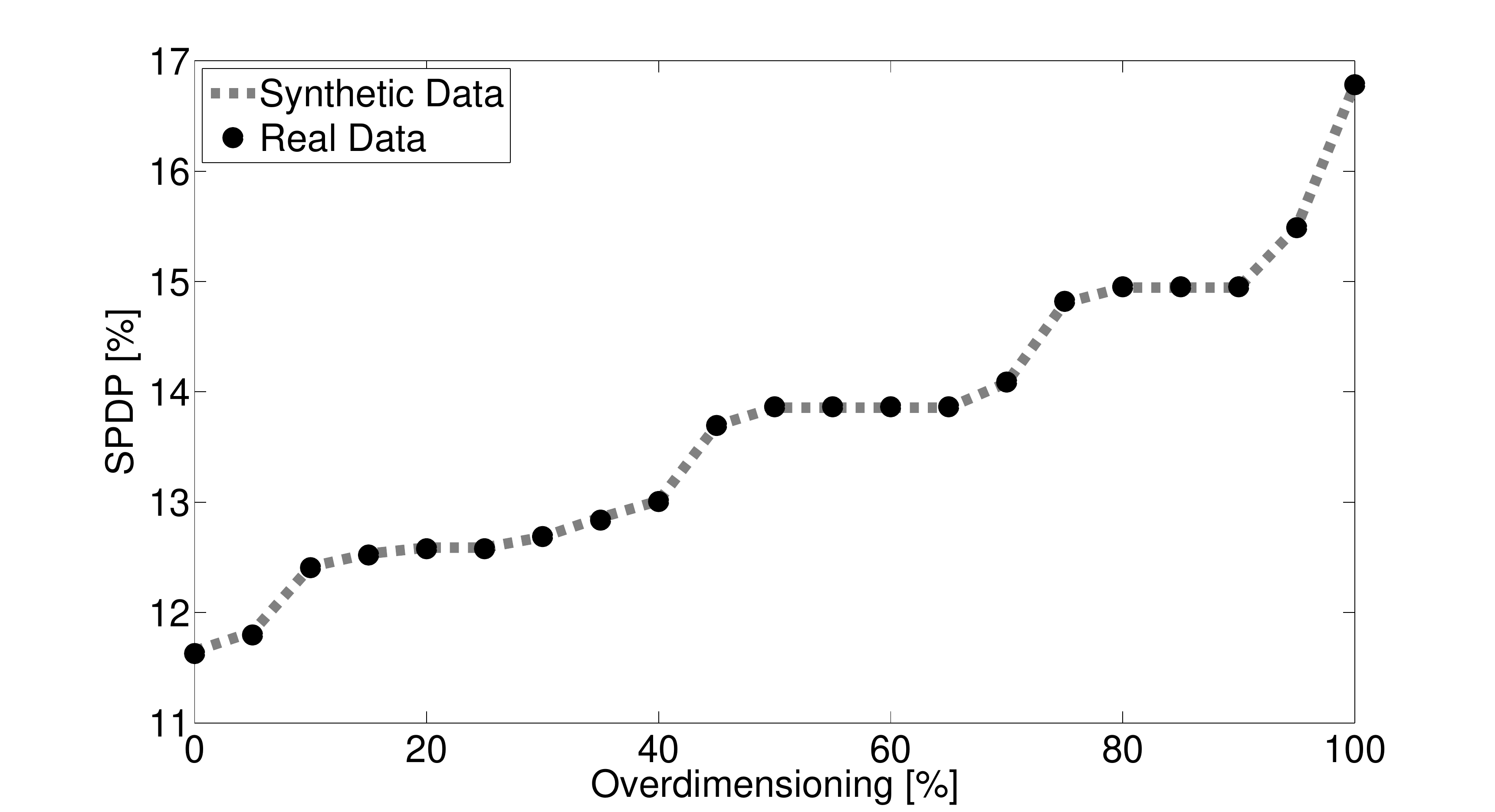}
\caption{Variation of the SPDP for a 5 years time period as a function of the rated power overdimensioning}\label{spdp}
\end{figure}

In Figure \ref{spdp} we show the comparison of the SPDP evaluated on real and synthetic data for a period of five years. Moreover we compute it by varying the over-dimensioning of the plant. In other words, we fix the maximum power required during the day at 9.8 $kW$ and we suppose to oversize the rated power of the wind turbine from 0 to 100 \% of its value. For example if we have a wind turbine with a rated power of 10 $kW$ and we want to improve the plant, we are interested in knowing how the SPDP will change. If we assume a 40\% of over-dimension (then we will have a turbine of 10+4 $kW$), the SPDP will be 13\%, with a percentage variation of about 10\%. 

It is interesting to note that the variation of the SPDP is very important, it increases almost of 50\% with a variation of 100\% of the overdimensioning. This behavior is well kept by the model.

\subsection{Financial indicators}

In this subsection we will compare important financial indicators computed on real time series, generated through the bootstrapping procedure, and simulated data, generated through the ISMC model by means of Monte Carlo simulations.We consider two different cases. In the first case we compute the cash flow obtained by selling the energy produced by the wind turbine taking into account the Italian government incentives on the selling price. In Italy for a wind farm with a rated power under 1 $MW$ and until 15 years from the setup of the plant, the price of the energy is fixed, at the time of writing, to 0.291 \euro$/kWh$. In this case we consider a cash flow of 15 years. Just to give a quantitative example, the second case takes into account the selling of energy at the present market price in Italy, without the government incentives, which is 0.027 \euro$/kWh$. In this case the considered period of the cash flow is extended to 30 years.

For each case we generate 1000 possible scenarios and we evaluate the financial indicators for both real and synthetic data. The trajectories are transformed into power through the power curve of the wind turbine chosen, then in energy by considering the time between two successive steps and after in euro by multiplying for the specific price considered. The annual cash flow are built and at time $0$ the cost of the initial investment is considered. Commonly for wind turbines with a rated power under 100 $kW$ it is considered a price of 3000 \euro $\;$ for each $kW$ (this value comes from some quotes requested to sector companies). This price takes into account the cost of the turbine, the setup cost and also the annual maintenance cost. Given that our wind turbine a 10 $kW$ of rated power, we choose the initial investment of 30000 \euro.\\
\indent The considered indicators are the Semi-Elasticity and the Relative Convexity. We show the histograms of the financial indicators evaluated for all the 1000 scenarios, for real and simulated data, and for the two cases with and without the government incentives.

\subsubsection{Indexes for the measurement of the interest rate risk}

\begin{figure}
\centering
\subfloat[]{\label{main:sa}\includegraphics[height=7cm]{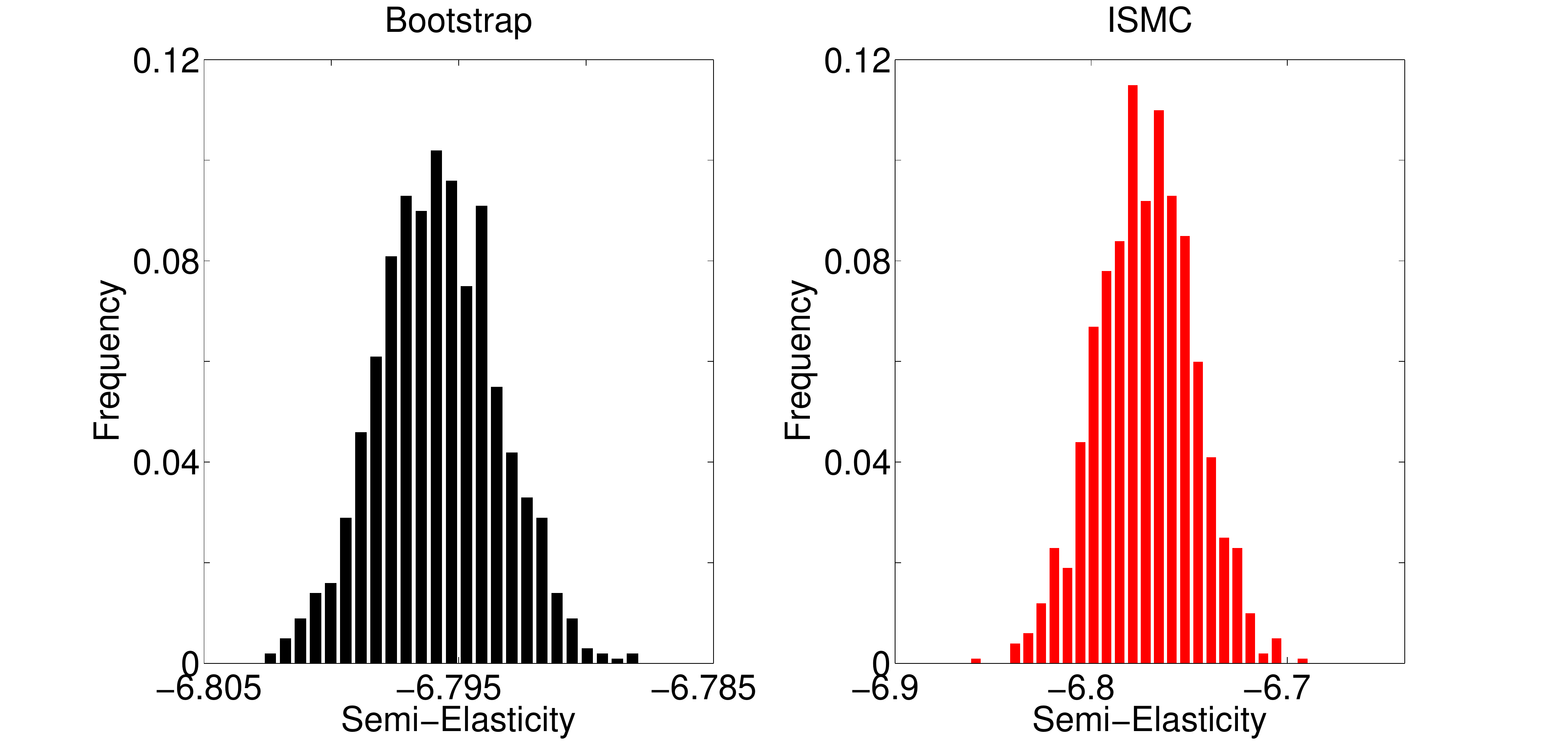}}

\subfloat[]{\label{main:cs}\includegraphics[height=7cm]{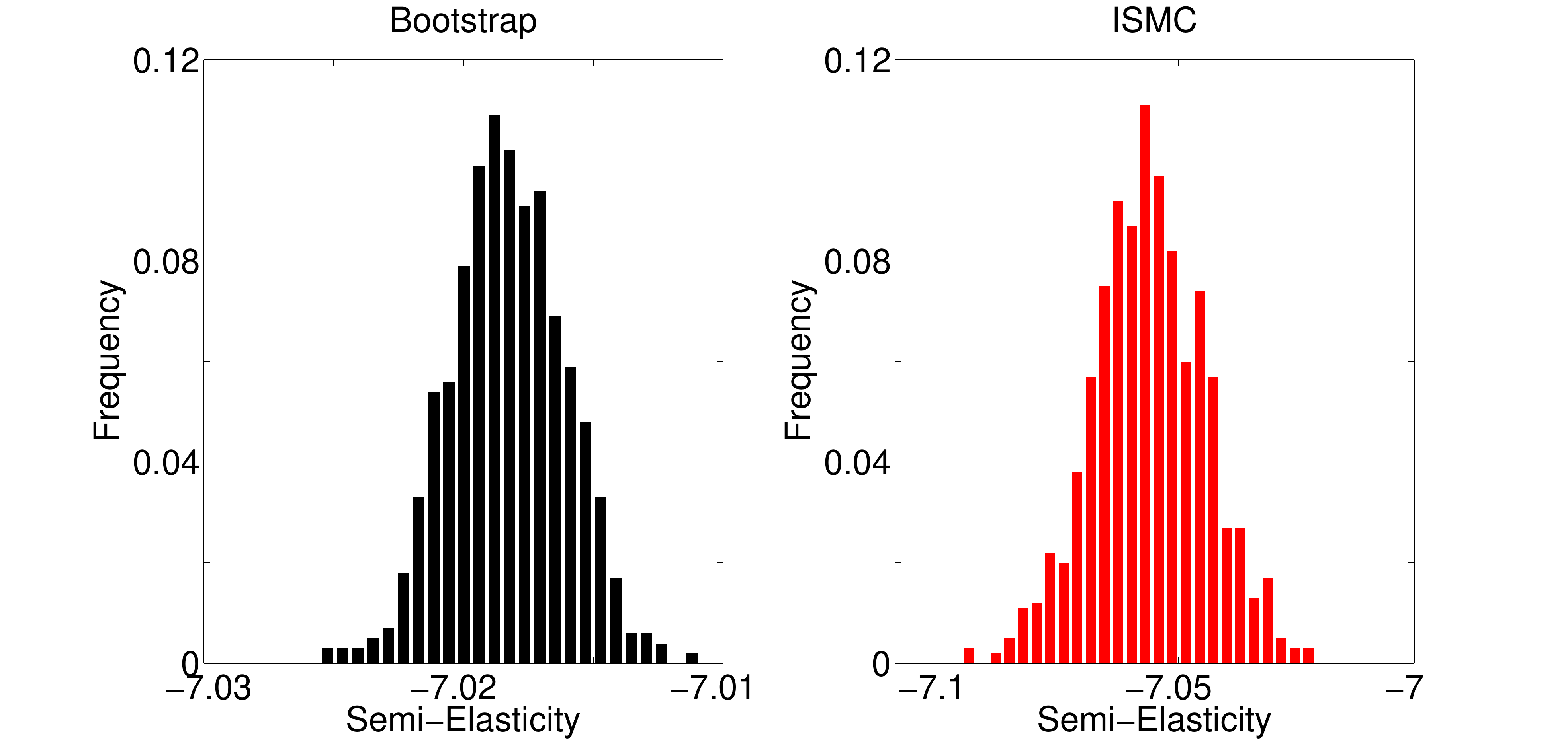}}
\caption{Histograms of the Semi-elasticity for real and simulated data. (a) Case with government incentives and cash flow of 15 years. (b) Case without government incentives and cash flow of 30 years.}
\label{hist2}
\end{figure}

In this subsection we evaluate some measures of the interest rate risk that are commonly used in financial analysis and we apply them to the investment in a wind farm. As we have seen, the production of energy generates a cash flow sequence. The value of this stream may change in time due to variations of the interest rate. The wind energy producer, as any other firm, may suffer the risk of changes in the interest rate and then it is necessary to control such kind of uncertainty.\\
\indent One of the most important index is the Duration ($D$) that was introduced by Mcaulay (1938). It is defined as a weighted average of time to payment of the cash flow sequence. From the Duration it is possible to recover the semi-elasticity ($Se$) that is a relative measure of the variation of the value of the cash flow stream. In formula we have:
\begin{equation}
Se=-\frac{1}{1+r}\frac{\sum_{s=1}^{n} s \cdot CF_s \cdot (1+r)^{-s} }{ \sum_{s=1}^{n} CF_s \cdot (1+r)^{-s}}=-\frac{D}{1+r}.
\end{equation}
In the application the interest rate $r$ was equal to 3\% and the results on the semi-elasticity are summarized in Figure \ref{hist2}. In Table \ref{t2} instead, we give some numeric values of the statistics of the distributions.

\begin{table}
\begin{center}
\begin{tabular}{lccccccc}
\hline
Semi-Elasticity        &        &        &          &       & & &   \\
\hline
          		& Mean   & STD    & Skewness & Kurtosis & JB & jB stat & p-value \\
Real 15   		& -6.79 &    0.0024 &   0.0758  &  2.92 &    Accepted &    1.21 &   0.50  \\
Synthetic 15    & -6.77 &    0.0246 &  -0.0658  &  3.01 & Accepted &  0.72  &  0.50  \\
Real 30   		& -7.03  &  0.0022 &   -0.0183 &   3.03  &  Accepted &    0.11  &  0.50 \\
Synthetic 30    & -7.05 &    0.0118 &  -0.0110 &   3.10  &   Accepted &    0.43 &    0.50  \\
\hline
\end{tabular}
\caption{Statistics of semi-Elasticity distributions evaluated for real and synthetic data and for the two cases: with government incentives (15 years) and without (30 years).}
\label{t2} 
\end{center}
\end{table}

Another important index is the Relative Convexity (RC) that expresses the convexity of the discounted value of the cash flow stream in unit of value variation. In formula we have:
\begin{equation}
RC=\frac{1}{Se}  \frac{\sum_{s=1}^{n} s \cdot (s+1) \cdot CF_s \cdot (1+r)^{-s} }{ \sum_{s=1}^{n} CF_s \cdot (1+r)^{-s}}=\frac{C}{Se}
\end{equation}
and $C$ denotes the convexity of the cash flow. The behavior of the duration and of the convexity were computed, by the same authors, in \cite{windbar}.

\begin{figure}
\centering
\subfloat[]{\label{main:a}\includegraphics[height=7cm]{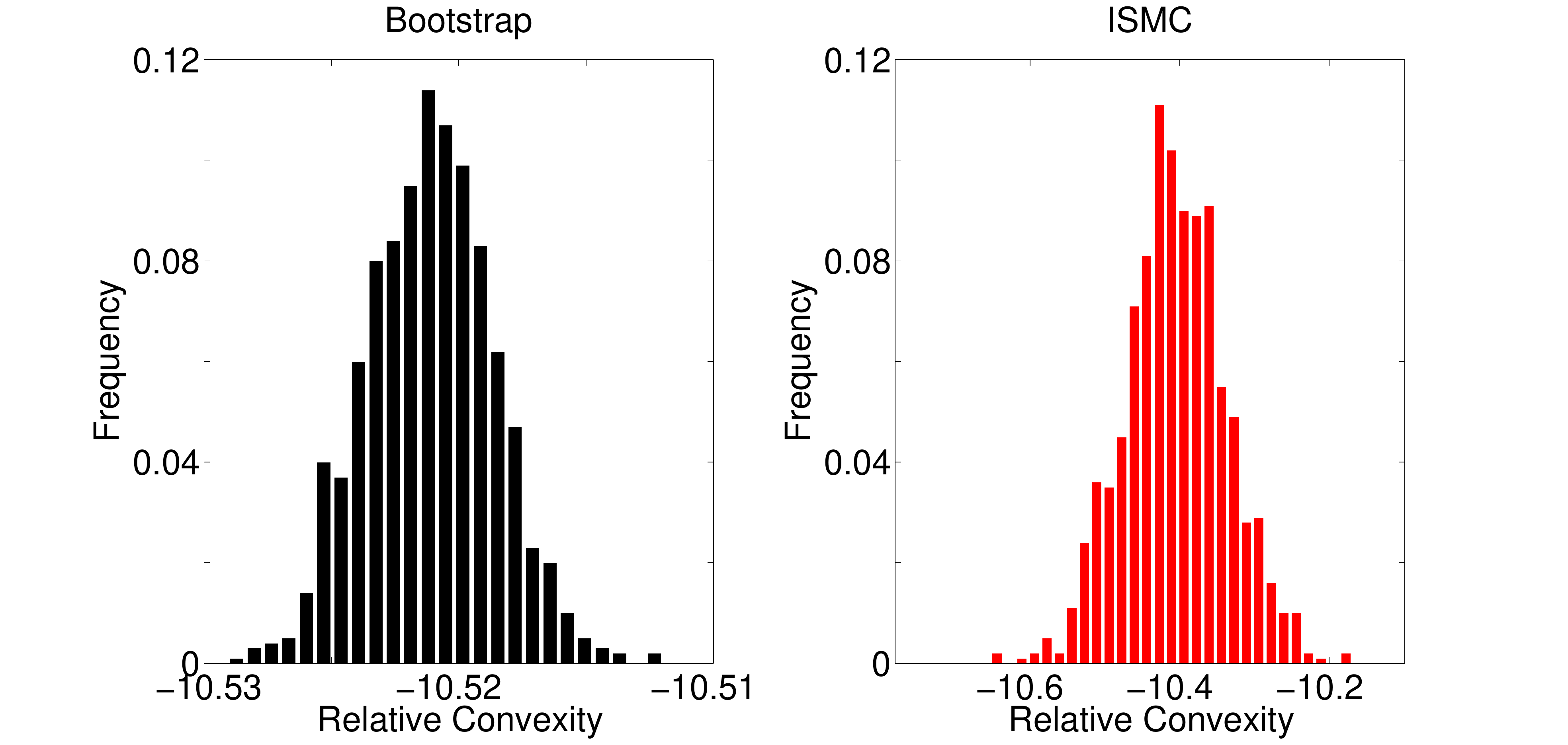}}

\subfloat[]{\label{main:c}\includegraphics[height=7cm]{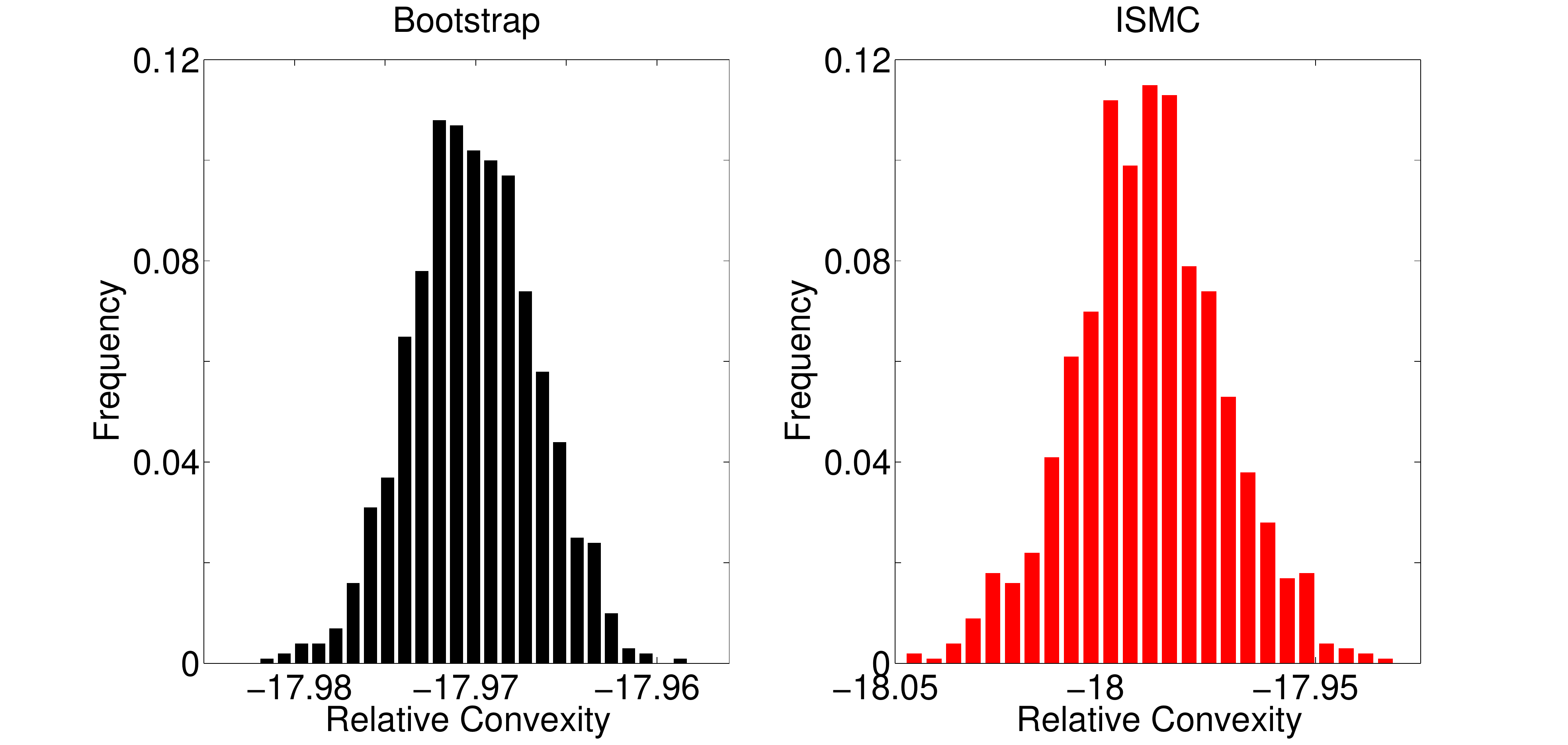}}
\caption{Histograms of the relative Convexity for real and simulated data. (a) Case with government incentives and cash flow of 15 years. (b) Case without government incentives and cash flow of 30 years.}
\label{hist3}
\end{figure}

In Figure \ref{hist3} is plotted the Relative Convexity in the same cases of the previous indicators and in Table \ref{t3} there are its statistic values.

\begin{table}
	\begin{center}
		\begin{tabular}{lccccccc}
			\hline
			Relative Convexity        &        &        &          &       & & &   \\
			\hline
			& Mean   & STD    & Skewness & Kurtosis & JB & jB stat & p-value \\
			Real 15   		& -10.52 &  0.0025 &   0.0768 &   3.07  & Accepted &  1.2277  &  0.50 \\
			Synthetic 15    & -10.40 &   0.0677 &  0.0226  &  3.24 & Accepted  & 2.6253  &  0.50  \\
			Real 30   		& -17.97 &   0.0035 &  -0.0024  &  2.99   & Accepted & 0.0013 &   0.50 \\
			Synthetic 30    & -17.98 &  0.0176 &    0.0050 &   3.16 & Accepted & 1.1506  &  0.50  \\
			\hline
		\end{tabular}
		\caption{Statistics of Relative Convexity distributions evaluated for real and synthetic data and for the two cases: with government incentives (15 years) and without (30 years).}
		\label{t3} 
	\end{center}
\end{table}

As it is possible to note, for each pair of indicators and for each specific case of the period of investment, the mean values of the indicators are almost the same for real and simulated data.

%
%

\section{Conclusion} 

In this paper we apply an indexed semi-Markov model (ISMC) to compute different financial indicators like the Internal Rate of Return, the Semi-Elasticity and the Relative Convexity. To do this we use a real scaled wind speed database and a commercial wind turbine in order to compute the real energy produced. To conduce an economic analysis on a medium therm investment on a wind turbine, we apply a bootstrap procedure. Moreover, we propose a new indicator that can quantify the degree of utilization of a power plant. By means of Monte Carlo simulation we show that the ISMC can reproduce almost exactly all the indicators calculated on real data.

\bibliographystyle{prattico_bib} 
\bibliography{b}

\end{document}